\documentclass[12pt]{article}
\usepackage[margin=2cm]{geometry}
\usepackage[latin1]{inputenc}
\usepackage{mathrsfs}
\usepackage{amsfonts}
\usepackage{amstext}
\usepackage{amsmath}
\usepackage{amssymb}
\usepackage{graphics,graphicx}
\usepackage[font=small,labelfont=bf]{caption}
\usepackage{xcolor}
\usepackage{upgreek}

\newcommand{\hi } {{\rm H}\,{\small\rm I}}
\newcommand{\hii } {{\rm H}\,{\small\rm II}}

\title{\textbf{Gas dynamics in dwarf galaxies as testbeds for dark matter and galaxy evolution}}

\author{Federico Lelli \\ 
\normalsize{INAF, Arcetri Astrophysical Observatory, Florence 50125, Italy;}\\ 
\normalsize{E-mail: federico.lelli@inaf.it}}

\date{}

\begin{document}

\maketitle

\textbf{Dwarf galaxies are ideal laboratories to test dark matter models and alternative theories because their dynamical mass (from observed kinematics) largely outweighs their baryonic mass (from gas and stars). In most star-forming dwarfs, cold atomic gas forms regularly rotating disks extending beyond the stellar component, thus probing the gravitational potential out to the outermost regions. Here I review several aspects of gas dynamics in dwarf galaxies, such as rotation curves, mass models, and noncircular motions. Star-forming dwarfs extend the dynamical laws of spiral galaxies to lower masses, surface densities, and accelerations. The three main dynamical laws of rotation-supported galaxies point to three distinct acceleration scales, which play different physical roles but display the same value, within uncertainties. The small scatter around these dynamical laws implies a tight coupling between baryons and dark matter in galaxies, which will be better understood with next-generation surveys that will enlarge current sample sizes by orders of magnitude.}

\section{Introduction}

Gas dynamics plays a key role in the formation and evolution of galaxies. During the history of our Universe, cosmic gas accreated onto galaxies, cooled down, and settled into regularly rotating disks where new stars are formed. Gas disks are precious tools to trace the gravitational potential ($\Phi$) of galaxies because (unlike stellar disks) the rotation support largely dominates over the pressure support in the radial direction ($R$), thus the observed rotation velocity is virtually equal to the circular velocity of a test particle ($V_{c}^2 = -R\,\nabla \Phi$). For example, the discovery of ubiquitous flat rotation curves in spiral galaxies using H$\alpha$ \cite{Rubin1978} and \hi\ \cite{Bosma1978} emission-line observations was pivotal to establish the dark matter (DM) problem. 

In high-surface-brightness (HSB) and high-mass galaxies (with stellar masses $M_\star > 3 \times 10^{9}$\,M$_\odot$), the visible baryonic matter can usually explain the inner parts of the observed rotation curves, such as those from H$\alpha$ observations \cite{Kalnajs1983, Kent1986}, while the DM effect becomes evident in the outer regions, which can be traced with \hi\ observations \cite{vanAlbada1985, Kent1987}. In low-surface-brightness (LSB) and low-mass galaxies ($M_\star \leq 3 \times 10^{9}$ M$_\odot$), instead, the DM effect is already important at small radii \cite{Carignan1988a, deBlok1997, Mateo1998}. Thus, LSB and low-mass galaxies have been prime targets to test DM models because the modeling of the baryonic mass distribution is less crucial than in HSB and high-mass galaxies. In addition, dwarf galaxies allow extending the study of dynamical laws down to the lowest accessible masses, surface densities, and accelerations, where DM is expected to play a major role in galaxy evolution according to the $\Lambda$ Cold Dark Matter ($\Lambda$CDM) cosmological model. Dwarf galaxies pose several challenges to the $\Lambda$CDM paradigm, which have been extensively reviewed \cite{deBlok2010, Bullock2017, Pawlowski2018}. This review is focused on gas dynamics in dwarf galaxies from an empirical perspective, and discusses only the small-scale $\Lambda$CDM challenges posed by gas emission-line observations.

Over the past 20 years, enormous progress has been made in our understanding of gas dynamics in dwarf galaxies thanks to dedicated surveys using \hi\ interferometry, H$\alpha$ Fabry-P\'erot observations, and optical integral-field spectroscopy. In my view, the key advances have been (i) the confirmation that dwarf galaxies have slowly rising rotation curves in their inner regions, implying the existence of DM cores in a $\Lambda$CDM context \cite{deBlok2010}, (ii) the discovery that gas-dominated dwarfs fall on the same baryonic Tully-Fisher relation (BTFR) as star-dominated spirals, posing a fine-tuning problem for $\Lambda$CDM models of galaxy formation \cite{McGaugh2000, McGaugh2012}, (iii) the establishment of new dynamical laws such as the central density relation (CDR) \cite{Swaters2014, Lelli2016c} and the radial acceleration relation (RAR) \cite{McGaugh2004, McGaugh2016, Lelli2017}, extending from high-mass to low-mass galaxies. In particular, the main dynamical laws of rotation-supported galaxies (BTFR, CDR, and RAR) have been found to have small intrinsic scatters, implying that baryons and DM must be tightly coupled on sub-kpc scales. This review will summarize this progress but also highlight the key scientific questions that remain to be addressed: (i) Are the observed dynamical properties of dwarf galaxies (rotation curve shapes, DM content, strength of noncircular motions) consistent with the expectations from $\Lambda$CDM models of galaxy formation? 
(ii) Do the properties of the dynamical laws depend on the galaxy large-scale environment? (iii) Is there any intrinsic scatter along the dynamical laws, especially in the dwarf regime where DM is expected to largely dominate the dynamics? 

\subsection{Taxonomy of dwarf galaxies}

The dwarf galaxy classification covers more than 7 orders of magnitude in stellar mass, from a few billions to a few hundreds solar masses \cite{Bullock2017}. This mass range is much broader than the $\sim$2.5 dex covered by ellipticals, lenticulars, and spirals along the Hubble sequence. This review will focus on the mass range $M_\star \simeq 10^{5.5}-10^{9.5}$ M$_\odot$ where we find both gas-rich dwarfs (in which we can simultaneously study gas and stellar dynamics) and gas-poor dwarfs (in which we are limited to stellar dynamics). I will not discuss the ultra-faint dwarfs (UFDs) with $M_\star < 10^{5.5}$ M$_\odot$ found in the Local Group because they are generally devoid of detectable gas \cite{Tolstoy2009}. Next-generation optical and \hi\ surveys, however, may potentially discover gas-rich and/or star-forming UFDs in the field, providing key constraint to $\Lambda$CDM models down to the lowest possible masses \cite{Bullock2017}.

The taxonomy of dwarf galaxies is complex due to historical reasons \cite{Binggeli1994}. I will use an extremely simplified nomenclature that distinguish two main types of dwarfs based on their gas content and star-formation activity:
\begin{itemize}
 \item \emph{Quiescent dwarfs} have no recent (the past 10 Myr) star formation \cite{Belokurov2021} and are generally undetected by \hi\ surveys (``gas-poor dwarfs''). This definition includes dwarf spheroidals, dwarf ellipticals, and dwarf lenticulars. Quiescent dwarfs are copiously found in galaxy clusters \cite{Ferguson1994} and as satellites of high-mass galaxies in groups \cite{Binggeli1990, Putman2021} but are extremely rare in isolation \cite{Binggeli1990, Geha2012}. 
 \item  \emph{Star-forming dwarfs} form stars at a steady rate \cite{Tosi2021} and are usually detected by \hi\ surveys (``gas-rich dwarfs''). This definition encompasses dwarf irregulars, star-forming LSB dwarfs, and late morphological types such as Sd, Sm, and Im galaxies. Star-forming dwarfs are found in isolation, in galaxy groups, and in the outskirts of galaxy clusters \cite{Binggeli1990, Kovac2009}.
\end{itemize}
In addition to these two main types of low-mass galaxies, I will occasionally refer to more exotic gas-rich dwarfs:
\begin{itemize}
 \item \emph{Starburst dwarfs} have enhanced star-formation rates (SFRs). Various definitions of starbursts exist in the literature; an effective one for dwarf galaxies is that the recent SFR is at least 3 times higher than the past average SFR \cite{McQuinn2010a, Lelli2014a}. This definition results in typical burst durations of a few hundreds Myrs \cite{McQuinn2010b, McQuinn2012}. Starburst dwarfs include blue compact dwarfs and other historical nomenclatures such as amorphous galaxies \cite{Gallagher1987} and \hii\ galaxies \cite{Terlevich1991}.
 \item \emph{Transition dwarfs} have some gas but no current star formation \cite{Mateo1998, Skillman2003}. They are usually found among the least massive gas-rich dwarfs and may represent the evolutionary link between star-forming and quiescent dwarfs \cite{Tolstoy2009}.
 \item \emph{Tidal dwarf galaxies} (TDGs) are recycled objects that form out of collisional debris ejected from more massive galaxies during galaxy interactions \cite{Duc1998, Lelli2015}. It is unclear how many dwarf galaxies may have a tidal origin with profund implications for $\Lambda$CDM cosmology \cite{Bournaud2006, Kroupa2012}.
\end{itemize}

I will not distinguish between dwarf galaxies and so-called ultra-diffuse galaxies (UDGs) because there is compelling evidence that the latter are a LSB extension of the former \cite{Conselice2018, Chilingarian2019, Brook2021}, including both gas-rich and gas-poor objects \cite{Karunakaran2020}.

\subsection{The multiphase gas content of dwarf galaxies}

The inter-stellar medium (ISM) of star-forming dwarfs is mostly composed of cold atomic gas ($T\simeq10^{2}-10^{4}$ K), which is directly traced by the \hi\ emission line at 21 cm. This is a hyperfine transition of atomic hydrogen, for which the \hi\ luminosity can be readily converted into a \hi\ mass (minor corrections for \hi\ self-absorption may possibly be necessary in edge-on disks \cite{Peters2017}). The atomic gas mass can then be estimated using statistical corrections for the minor contributions of helium and heavier elements \cite{McGaugh2020}. 

Cold molecular gas ($T\simeq10$ K) has been historically difficult to study in dwarf galaxies because they have low gas volume densities and low gas metallicities, so CO lines (tracing H$_2$) are hard to detect \cite{Schombert1990, Taylor1998}. Even when CO is detected, the molecular gas mass remains uncertain due to the uncertain CO-to-H$_2$ conversion factor \cite{Bolatto2013}. Indirect estimates of the H$_2$ mass, extrapolating the scaling relations with $M_\star$ and/or SFR of high-mass galaxies, suggest that molecules make between 3\% and 10\% of the ISM mass of star-forming dwarfs \cite{McGaugh2020, McGaugh2017, Hunt2020}. Recent CO observations with the Atacama Large Millimeter/submillimeter Array (ALMA) have been able to detect giant molecular clouds in low-mass galaxies, including low-metalliticy dwarfs \cite{Rubio2015}, starburst dwarfs \cite{Miura2018}, and TDGs \cite{Querejeta2021}. These detections represent major progress for understanding star formation in dwarf galaxies, but CO emission does not seem to be a promising kinematic tracer because its distribution is irregular and often made of distinct CO clumps at random galaxy locations.

Warm ionized gas ($T\gtrsim10^{4}$ K), probed by optical recombination lines (e.g., the H$\alpha$ line), provides an insignificant contribution to the ISM mass \cite{Kennicutt1984}. Still, it is a precious kinematic tracer in the inner galaxy regions because optical observations generally reach higher spatial resolutions than radio \hi\ interferometry \cite{Relatores2019, Barat2020}. H$\alpha$ and \hi\ lines are the most widely used kinematic tracers in star-forming dwarfs.

Hot ionized gas ($T\gtrsim10^5-10^6$ K) is difficult to probe in dwarf galaxies because one expects weak, diffuse X-ray emission. $\Lambda$CDM models of galaxy formation predict that low-mass galaxies should be surrounded by large amounts of hot gas, containing a significant fraction of the cosmological missing baryons \cite{Dutton2017, Katz2018}. Observationally, diffuse X-ray emission has been detected around a few starburst dwarfs but the visible hot gas mass is $\sim$1$\%$ or less than the \hi\ mass \cite{Ott2005}.

\begin{figure}
\includegraphics[width=0.97\linewidth]{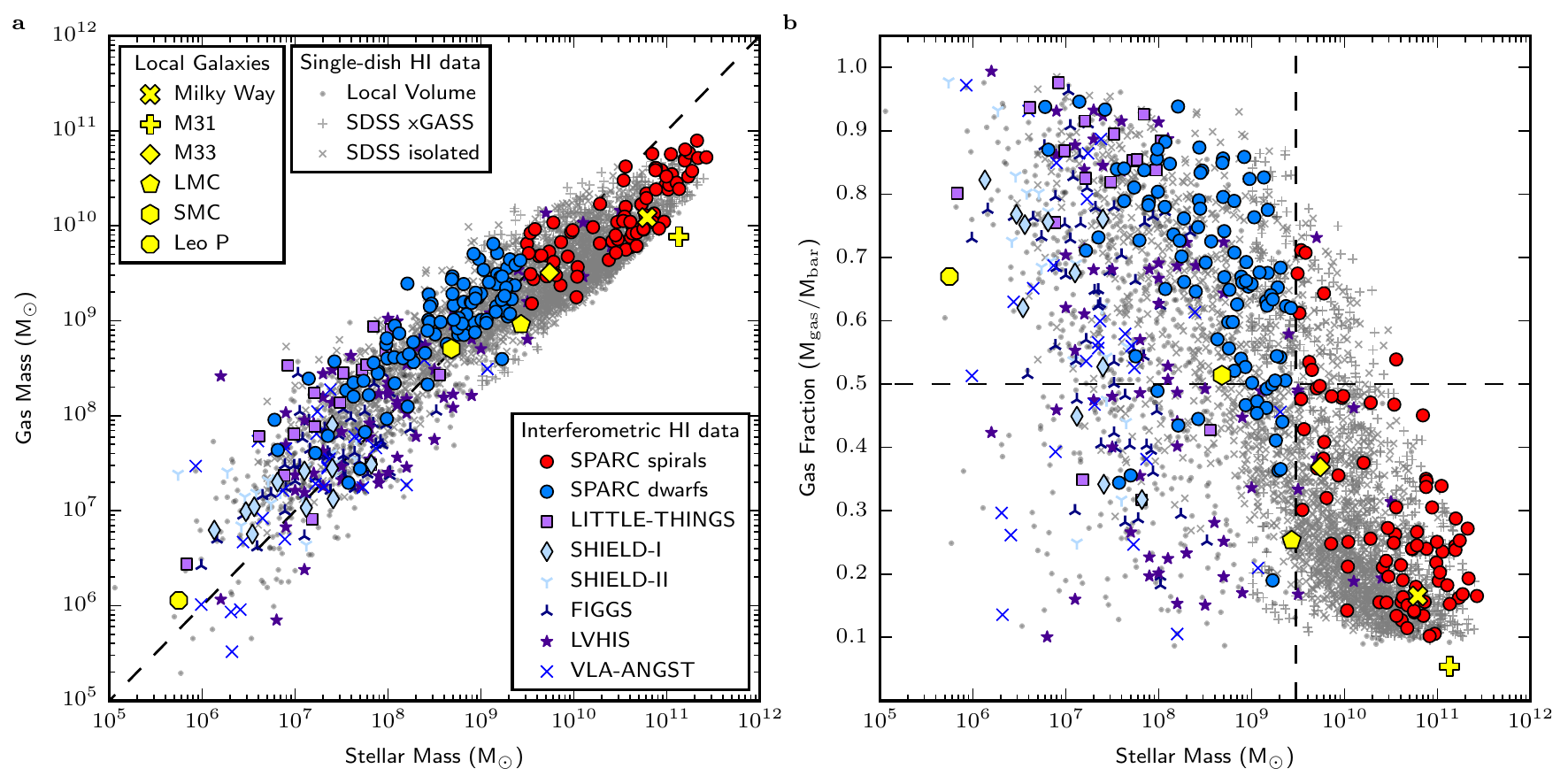}
\caption{\textbf{The cold gas content of nearby galaxies.} Gas mass (a) and gas fraction (b) are plotted as a function of stellar mass for \hi-detected galaxies. Yellow symbols show representative galaxies in or near the Local Group: the Milky Way (large cross) \cite{McGaugh2019}, M31 (large plus) \cite{Corbelli2010}, M33 (large diamond) \cite{Corbelli2014, Kam2017}, the LMC (pentagon) \cite{Staveley2003, vanDerMarel2006}, the SMC (hexagon) \cite{DiTeodoro2019}, and Leo\,P (octagon) \cite{Bernstein2014, McQuinn2015}. Grey symbols show single-dish \hi\ data for Local Volume galaxies within 11 Mpc (small dots) \cite{Karachentsev2013}, high-mass SDSS galaxies (small plusses) \cite{Catinella2010}, and isolated SDSS galaxies (small crosses) \cite{Bradford2015}. The other colored symbols show interferometric \hi\ data from SPARC (circles) \cite{Lelli2016a}, LITTLE-THINGS (squares) \cite{Hunter2012, Iorio2017}, SHIELD-I (diamonds) \cite{McNichols2016}, SHIELD-II (tris down) \cite{McQuinn2021}, FIGGS (tris up) \cite{Begum2008}, LVHIS (stars) \cite{Koribalski2018} and VLA-ANGST (crosses) \cite{Ott2012}. Stellar masses are heterogeneous but mostly come from $K$-band or Spitzer [3.6] photometry using mass-to-light ratios of 0.65 and 0.50 M$_\odot$/L$_\odot$, respectively. Typical 1$\sigma$ uncertainties (not shown) are of the order of 30$\%$ on stellar masses and 10$\%$ on gas masses. In panel a, the dashed line shows the 1:1 line. In panel b, the horizontal line corresponds to $M_{\rm gas}>M_{\star}$ while the vertical line shows our definition of a dwarf galaxy ($M_{\star} \leq 3 \times 10^9 M_\odot$) set by the LMC stellar mass. }
\label{fig:GasFraction}
\end{figure}
In star-forming dwarfs, therefore, the mass contributions of cold molecular gas, warm ionized gas, and hot ionized gas are generally less than 10$\%$ of the atomic gas mass, thus smaller than the typical uncertainty in \hi\ fluxes. Figure\,1 shows the cold gas content of star-forming galaxies over $\sim$6 dex in stellar mass, where the cold gas mass $M_{\rm gas}$ is obtained from the observed \hi\ mass with statistical corrections for heavier elements and molecules \cite{McGaugh2020}. The baryon content of high-mass spirals is usually dominated by stars ($M_\star > M_{\rm gas}$) while that of low-mass dwarfs can be largely dominated by cold gas ($M_{\rm gas} > M_{\star}$), especially for isolated galaxies \cite{Bradford2015}. The gas fraction of star-forming dwarfs can be extremely high: up to $\sim$95$\%$ of the baryonic mass can be in gas rather than stars. This implies that (i) star-forming dwarfs convert atomic gas into stars in a very inefficient way; most likely the bottleneck is the conversion from \hi\ to H$_2$ \cite{Hunt2020}; (ii) star-forming dwarfs have extremely long gas depletion times ($\sim$10$-$100 Gyr); the existing \hi\ reservoir can amply sustain the SFR without the need of accreting external gas \cite{McGaugh2017}; (iii) star-forming dwarfs in isolation will not evolve into quiescent dwarfs in the foreseeable future \cite{Geha2012}.

\section{Gas dynamics}

\subsection{Gas distribution and kinematics}

Spatially resolved studies of gas dynamics in dwarf galaxies were pioneered in the 1980s and 1990s using \hi\ interferometry \cite{Carignan1988a}. During the past 20 years, the number of dwarf galaxies with spatially resolved data has increased significantly thanks to dedicated \hi\ surveys (see Fig.\,1), providing a broad overview of the gas distribution and kinematics in different types of dwarfs.

The vast majority of star-forming dwarfs with $M_\star\simeq10^{7.5}-10^{9.5}$ M$_\odot$ host rotating \hi\ disks, similarly to more massive spiral galaxies \cite{Begum2008, Swaters2002a}. Typical dwarf ``irregulars'' are very regular in their \hi\ emission \cite{Begum2008, Swaters2002a}: the optical morphology may look complex due to star-forming clumps but do not reflect the underlying gas kinematics. Star-forming dwarfs with lower masses ($M_\star\simeq10^{5.5}-10^{7.5}$ M$_\odot$) can host rotating \hi\ disks \cite{McNichols2016, McQuinn2021} but their rotation velocities ($V_{\rm rot}\lesssim 40$ km\,s$^{-1}$) become comparable to the typical \hi\ velocity dispersion ($\sigma_{\rm V}\simeq5-10$ km\,s$^{-1}$), so the gas disk cannot be entirely supported by rotation. If one aims to measure their circular velocities ($V_{\rm c}^2 = -R\,\nabla \Phi$), the observed rotation velocities need to be corrected for pressure support using the so-called asymmetric-drift correction \cite{Lelli2014a, Iorio2017}. In addition, small $V_{\rm rot}/\sigma_{\rm V}$ ratios imply that the gas disks cannot be ``razor thin'' as generally assumed when deriving rotation curves with 2-dimensional methods \cite{Oh2018}, so one needs to use 3-dimensional methods to model thick disks \cite{DiTeodoro2015, Kurapati2020} and possibly ``disk flares'', i.e., the outward increase of disk thickness with radius \cite{Bacchini2020}. Finally, in the tiniest star-forming dwarfs with $V_{\rm rot}\simeq10-15$ km\,s$^{-1}$, the rotation velocities become comparable to typical noncircular motions ($\sim$5 km\,s$^{-1}$ as discussed later), so the \hi\ kinematics is unavoidably more complex \cite{Bernstein2014}.

\begin{figure}
\includegraphics[width=\linewidth]{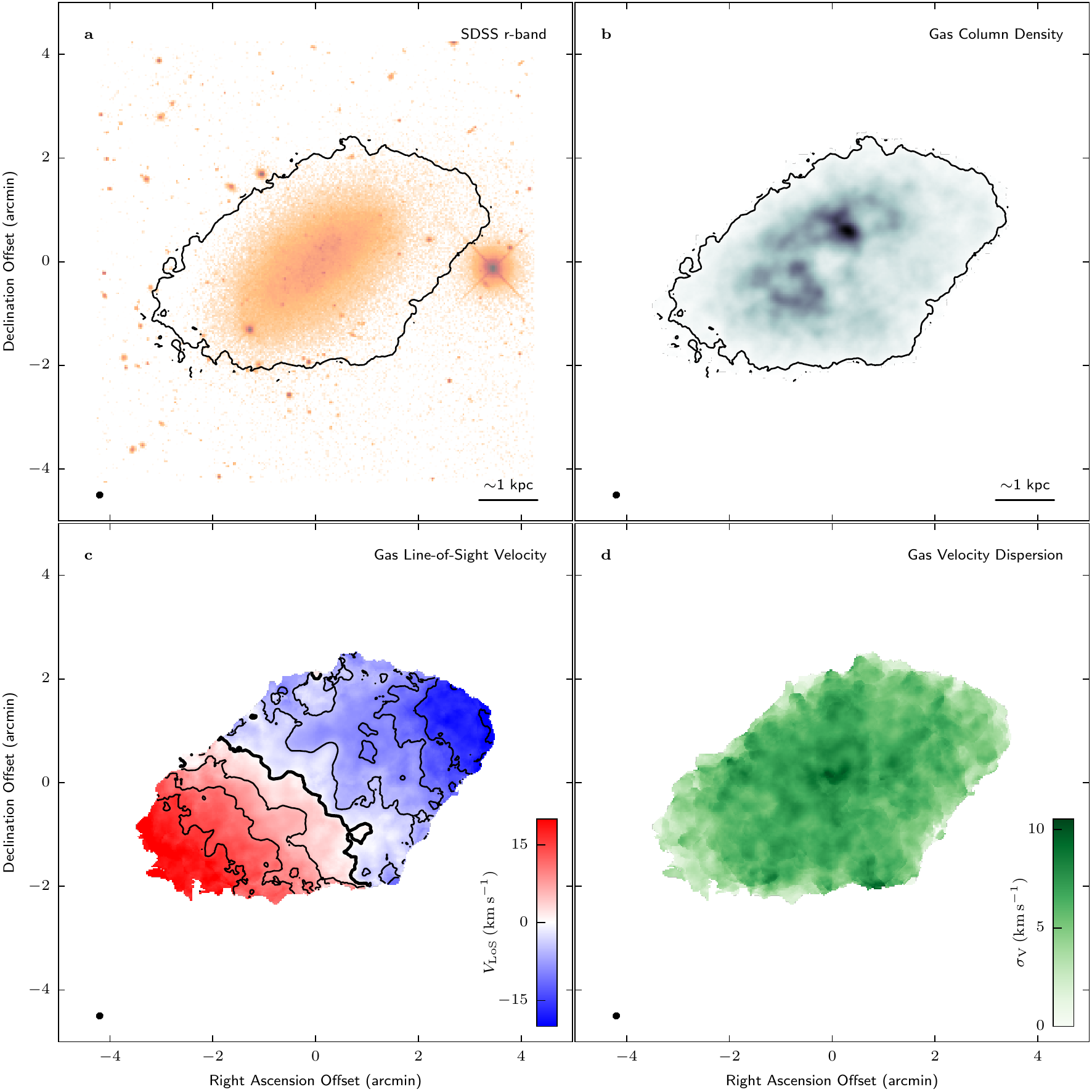}
\caption{\textbf{The star-forming dwarf galaxy DDO\,125}. The galaxy has a stellar mass of $\sim2 \times 10^7$\,M$_\odot$ \cite{Lelli2016a} and an outer rotation velocity of $\sim$20 km\,s$^{-1}$ \cite{Swaters2009}. The figure shows the $r-$band SDSS image (a), \hi\ column density map (b), \hi\ line-of-sight velocity map (c) and \hi\ velocity dispersion map (d). In panels a and b, the black contour corresponds to \hi\ column densities of $5\times10^{19}$ cm$^{-2}$, while the bar to the bottom-right corner corresponds to $\sim$1 kpc. In panels c and d, the colorbars show the velocity scale. In panel c, the bold contour corresponds to the systemic velocity (set to zero) while the other contours range from $-15$ to $+15$ km\,s$^{-1}$ in steps of 5 km\,s$^{-1}$. In all panels, the circle to the bottom-left corner shows the full-width half-maximum of the \hi\ point spread function \cite{Ott2012}.}
\label{fig:DDO125}
\end{figure}
Figure\,2 shows the example of the dwarf galaxy DDO\,125 \cite{Ott2012} with $M_\star\simeq2\times10^7$ M$_\odot$ \cite{Lelli2016a}. The \hi\ observations reveal a regularly rotating gas disk with a rotation velocity of only $\sim$20 km\,s$^{-1}$ in the outer parts and a velocity dispersion $\sigma_{\rm V}\lesssim10$ km\,s$^{-1}$. The \hi\ disk extends further out than the stellar component. On average, the ratio of \hi\ to stellar size in star-forming dwarfs is $1.8\pm0.8$ \cite{Swaters2002a}, similar to spiral galaxies in the field \cite{Broeils1997, Noordermeer2005}, with DDO\,125 being in the lower portion of the scatter. Extended \hi\ disks allow rotation curves to be traced beyond the stellar component, where the DM halo is expected to largely dominate the gravitational potential.

\begin{figure}
\includegraphics[width=\linewidth]{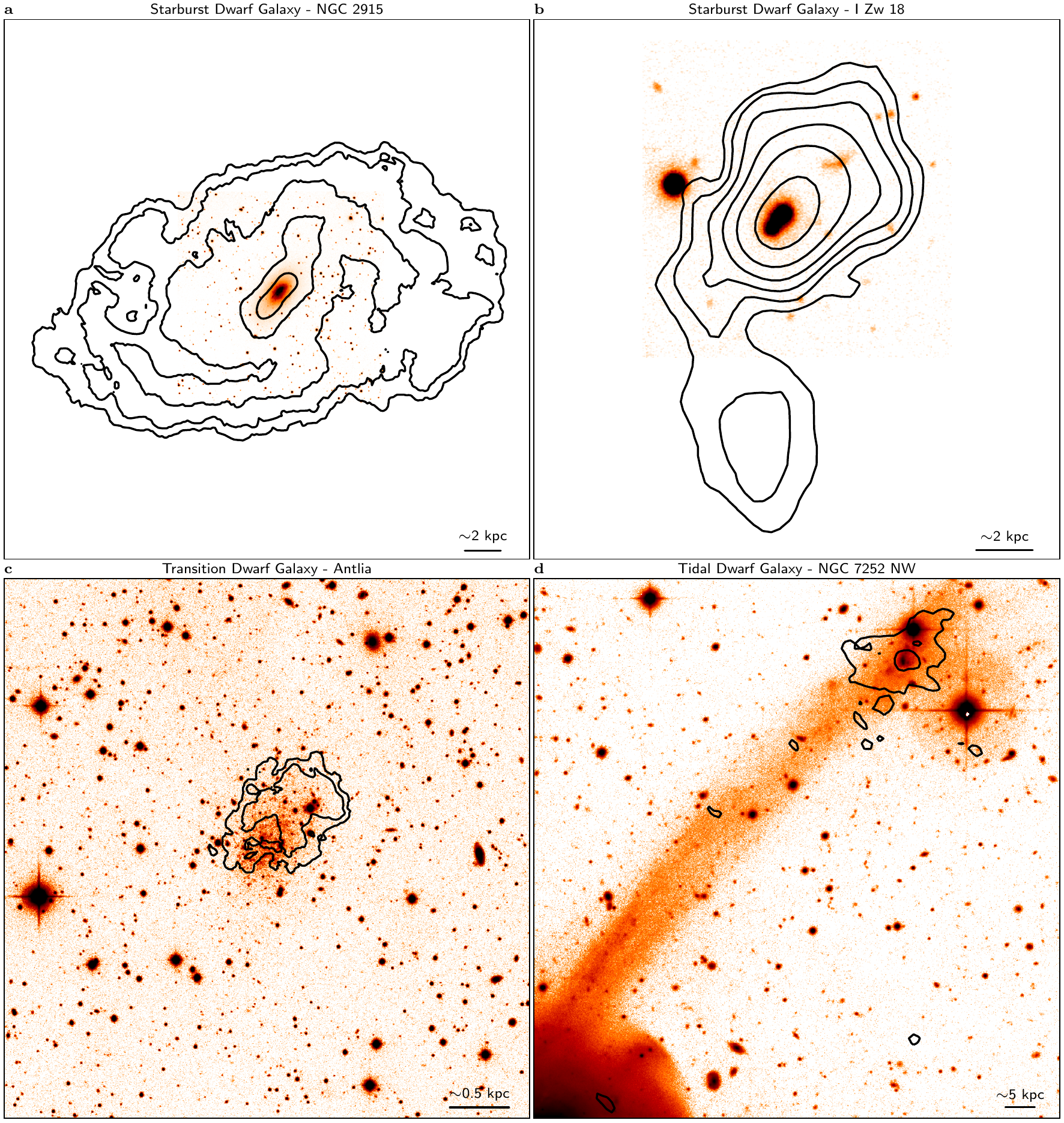}
\caption{\textbf{\hi\ distribution in peculiar types of dwarf galaxies.} Optical images (color scale) are overlaid with \hi\ column density maps (contours, corresponding to $\sim$3$\sigma$, 6$\sigma$, 12$\sigma$, and so on). In panel a, the starburst dwarf NGC\,2915 ($M_\star \simeq 3 \times 10^8$\,M$_\odot$, $M_{\rm gas}\simeq6\times10^8$\,M$_\odot$) shows a very extended \hi\ disk with spiral arms \cite{Elson2010}. In panel b, the starburst dwarf I\,Zw\,18 ($M_\star \simeq 0.9 \times 10^8$\,M$_\odot$, $M_{\rm gas}\simeq3\times 10^{8}$\,M$_\odot$) displays a long \hi\ tail to the South and a companion dwarf galaxy to the North \cite{Lelli2012a}. In panel c, the transition dwarf Antlia ($M_\star \simeq 7 \times 10^6$\,M$_\odot$, $M_{\rm gas}\simeq0.7\times 10^{6}$\,M$_\odot$) displays an off-center \hi\ distribution \cite{Ott2012}. In panel d, the TDG NGC\,7252\,NW ($M_\star \simeq 0.9 \times 10^8$\,M$_\odot$, $M_{\rm gas}\simeq7\times 10^{8}$\,M$_\odot$) is forming at the tip of a tidal tail around a late-stage major merger \cite{Lelli2015}. In all panels, the bar to the bottom-right corner shows the physical scale.}
\label{fig:HIex}
\end{figure}
Peculiar types of gas-rich dwarfs show more diversity. Starburst dwarfs, which are estimated to be $\sim$6$\%$ of local star-forming dwarfs \cite{Lee2009b}, can show either regularly rotating \hi\ disks (Fig.\,3, a) or disturbed \hi\ distributions and kinematics (Fig.\,3, b) with similar probability \cite{Lelli2014a}. The latter ones are likely due to dwarf-dwarf interactions \cite{Pearson2016, Lelli2014c}, gas accretion \cite{Ashley2017, LopezSanchez2012}, and/or stellar feedback \cite{Cannon2004, Kobulnicky2008}. Transition dwarfs can show unsettled \hi\ distributions with significant off-sets between stellar and \hi\ emission (Fig.\,3, c), suggesting either gas stripping from a star-forming dwarf or gas accretion into a quiescent dwarf \cite{Mateo1998, Koribalski2018}. TDGs, which form out of collisional debris, can host rotating \hi\ disks (Fig.\,3, d) but it is unclear whether they are in dynamical equilibrium because they did not have enough time to complete a single orbit since the time of their formation \cite{Lelli2015}.

Spectroscopic H$\alpha$ surveys are largely complementary to \hi\ surveys because they can probe the inner galaxy regions with higher spatial resolution \cite{Relatores2019}. The higher velocity dispersion of ionized gas ($\sim15-30$ km\,s$^{-1}$), however, pose an intrinsic limit in tracing H$\alpha$ circular velocities in galaxies with $M_\star\lesssim10^{8}$ M$_\odot$ \cite{Barat2020}: the pressure support becomes important and the asymmetric-drift correction uncertain. H$\alpha$ surveys are also useful to study gas outflows, as discussed in the section on noncircular motions.

\begin{figure}
\includegraphics[width=\linewidth]{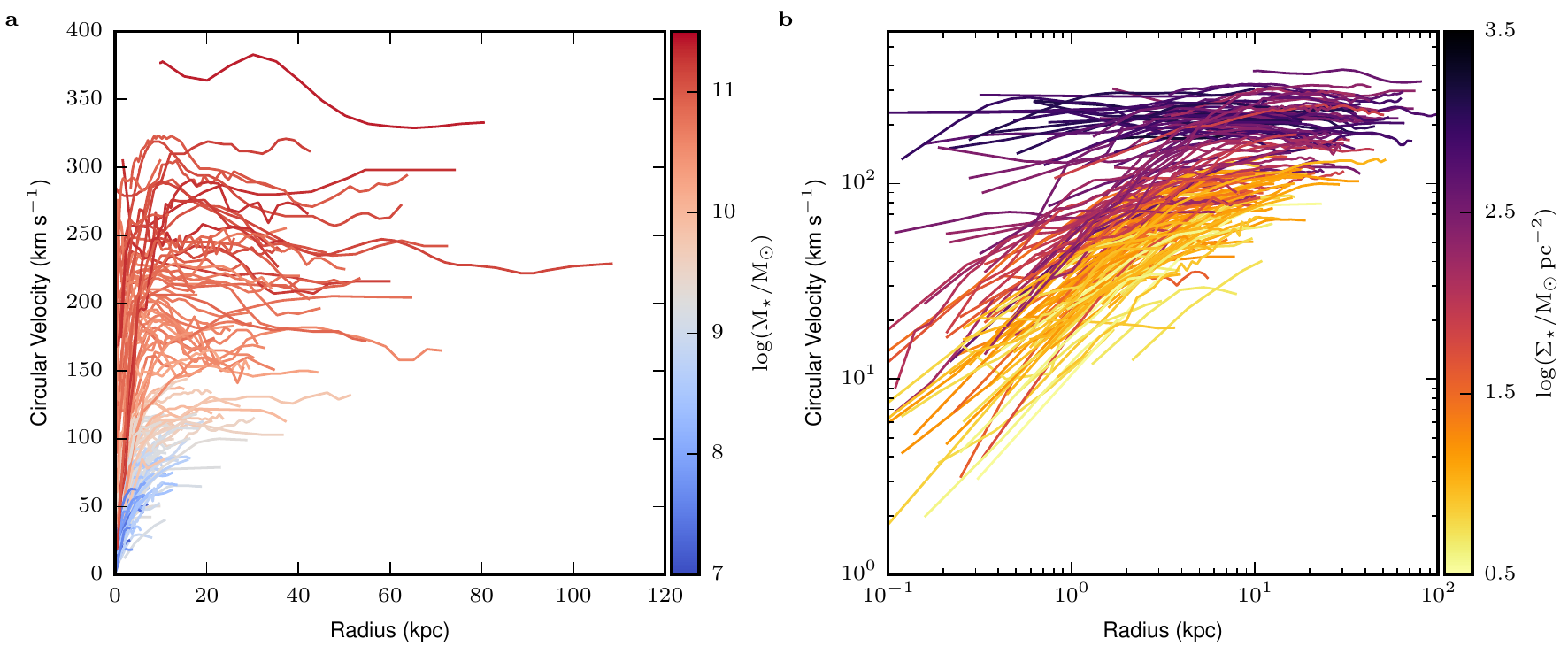}
\caption{\textbf{Rotation curves of dwarf and spiral galaxies from the SPARC database}. In panel a, rotation curves are plotted on a linear scale and color-coded by stellar mass. In panel b, they are plotted on a logaritmic scale and color-coded by effective stellar surface density. Typical 1$\sigma$ uncertainties on the rotation velocities (not shown) are of the order of 10$\%$, excluding inclination uncertainties that affect the overall normalization of the rotation curve \cite{Lelli2016a}.}
\label{fig:RCs}
\end{figure}
\subsection{Circular motions}

In spiral galaxies, the shapes of rotation curves have long been known to relate to the distribution of stellar light, suggesting that baryons dominate the gravitational potential in the inner regions \cite{Kalnajs1983, Kent1986, vanAlbada1985, Kent1987}. Remarkably, the same phenomenology holds in dwarf galaxies even though the DM halo is expected to dominate the gravitational potential down to the inner parts \cite{Swaters2009, Sancisi2004}. Figure\,4 shows hybrid H$\alpha$+\hi\ rotation curves for 175 disk galaxies from the SPARC database \cite{Lelli2016a}, spanning $\sim$5 dex in stellar mass. Panel a shows that the amplitude of the rotation curve systematically increases with stellar mass: this is the backbone of the Tully-Fisher relation. Panel b shows that the inner steepness of the rotation curve (i.e., how fast it reaches the flat part) systematically increases with stellar surface density: this is the backbone of the central density relation. Both relations are discussed in detail in the section on dynamical laws.

In a $\Lambda$CDM context, one may expect that the rotation curves of dwarf galaxies systematically vary with mass but show little variation at fixed mass, reflecting the expected self-similarity of the underlying DM halos \cite{Oman2015}. This is at odds with the observed rotation curves of dwarf galaxies: at fixed mass, HSB dwarfs have steeply rising rotation curves while LSB dwarfs have slowly rising rotation curves \cite{vanZee2001, Lelli2012b, Lelli2014b}. However, when the rotation curves are expressed in units of the disk scale length \cite{Swaters2009} or of the optical radius \cite{Karukes2017}, they do show a self-similar shape to a first-order approximation. The key problem is not the ``unexpected diversity'' but the remarkable regularity: at fixed mass, the shapes of rotation curves systematically depend on the baryon distribution (surface mass density and/or scale length), pointing to a local baryon-DM coupling.

When high-resolution and high-sensitivity data are available, rotation curves can display characteristic kpc-scale features that usually correspond to analogous features in the baryonic mass profiles \cite{Sancisi2004}. The best way to visualize this connection is to build mass models that account for the expected circular velocities from the main baryonic components: the gas and stellar disks (bulges are generally absent in dwarf galaxies). The gravitational contribution of a razor-thin exponential disk can be computed analytically \cite{Freeman1970} but this approximation neglects relevant features in the observed surface brightness profiles, such as bumps, wiggles, inner flattenings, and so on. Thus, it is more appropriate to numerically solve the Poisson's equation in cylindrical symmetry \cite{Casertano1983} for $\rho(R, z) = \Sigma(R)Z(z)$, where $\Sigma(R)$ is the observed radial density profile (derived from near-IR images for the stars and \hi\ maps for the gas) and $Z(z)$ is an adopted vertical density profile (e.g., an exponential or $sech^2$ distribution).

\begin{figure}
\begin{center}
\fbox{
 \begin{minipage}{\linewidth}
\textbf{Box 1 $-$ Mass models in a $\Lambda$CDM context}\newline
N-body simulations of structure formation in a $\Lambda$CDM cosmology find that DM halos follow the Navarro-Frenk-White (NFW) profile \cite{Navarro1996}, in which the volume density diverges to infinity towards the center (a so-called ``cusp''). The rotation curves of dwarf galaxies, instead, are best fitted with cored DM profiles, leading to the so-called ``cusp-vs-core'' problem \cite{deBlok2010}. A closely related issue is the ``too-big-too-fail'' problem, which considers the internal dynamics of dwarf galaxies (both quiescent and star-forming ones) concurrently with their observed cosmic abundance \cite{Papastergis2015, Papastergis2016, Read2017}. The figure shows a NFW fit to the dwarf galaxy NGC\,1560 \cite{Gentile2010}. The discrepancy at $R<3$ kpc illustrates the usual cusp-vs-core problem, which can be circumvented fitting cored DM halos \cite{Katz2017, Papastergis2017, Li2020} motivated by hydrodynamic simulations of galaxy formation with strong stellar feedback \cite{Governato2010, DiCintio2014, Read2016a}. The most interesting discrepancy, however, is the one concerning the entire rotation-curve shape. A smooth halo profile (NFW or any other form) cannot capture the characteristic features in the rotation curve. The actual DM density profile must also contain features mirroring those in the baryonic mass distribution.\\
\begin{center}
\includegraphics[width=0.5\linewidth]{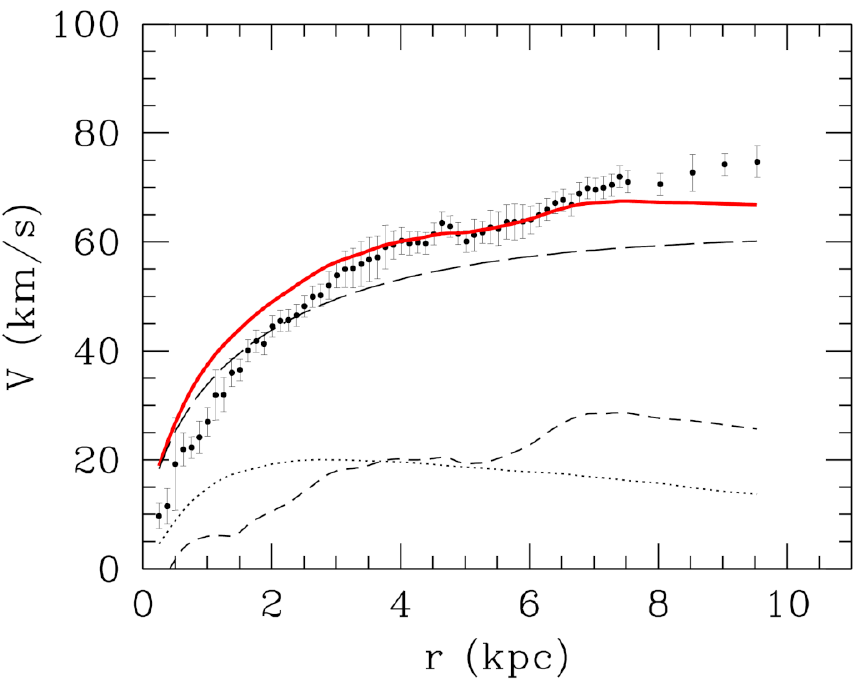}
\end{center}
\textbf{NFW fit to the dwarf galaxy NGC\,1560.} The observed rotation curve (dots with 1$\sigma$-errorbars) is compared to the best-fit mass model (red solid line) given by the gravitational contributions of stars (dotted line), gas (short-dashed line), and DM (long-dashed line). The galaxy distance is fixed to 3.4 Mpc \cite{Gentile2010}.
\end{minipage}
}
\end{center}
\end{figure}

\begin{figure}
\begin{center}
\fbox{
 \begin{minipage}{\textwidth}
\textbf{Box 2 $-$ Mass models in a Milgromian dynamics context}\newline
Milgromian dynamics (MOND) is an alternative to particle DM proposed in 1983 by Mordehai Milgrom \cite{Milgrom1983a, Milgrom1983b, Milgrom1983c}. MOND alters Newtonian gravity and/or inertia at accelerations smaller than $\sim$10$^{-10}$\,m\,s$^{-2}$, which are typical for the outer parts of spiral galaxies and for every radii in dwarf galaxies. MOND has been tested against hundreds of galaxy rotation curves \cite{Sanders2002, Famaey2012, Li2018}. The figure shows the example of the dwarf galaxy NGC\,1560 \cite{Gentile2010}. The rotation curve is well reproduced in MOND because its characteristic shape is mirrored by the characteristic shape of the gas contribution. In general, MOND made several successful predictions on galaxy scales but has problems on galaxy-cluster scales \cite{Famaey2012, Milgrom2014}. Relativistic extensions of MOND that can reproduce gravitational lensing, the speed of gravitational waves, the cosmic microwave background, and other cosmological observations are under active development \cite{Skordis2019, Skordis2021}.\newline
\begin{center}
\includegraphics[width=0.495\linewidth]{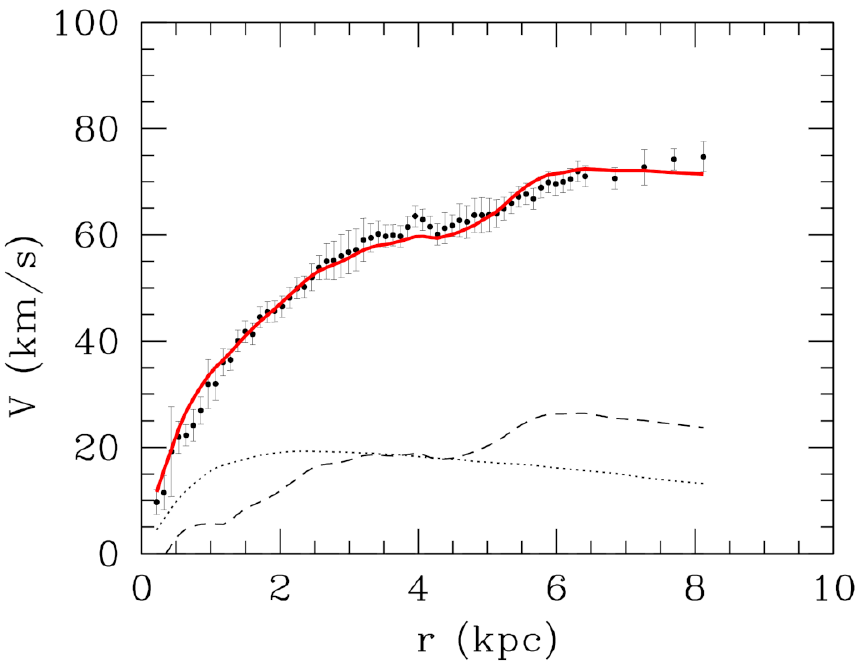}
\end{center}
\textbf{MOND fit to the dwarf galaxy NGC\,1560.} The observed rotation curve (dots with 1$\sigma$-errorbars) is compared with the MOND prediction (red solid line) based on the Newtonian contributions from stars (dotted line) and gas (dashed line). The best-fit galaxy distance is 2.9 Mpc \cite{Gentile2010}.
\end{minipage}
}
\end{center}
\end{figure}
The importance of such features is discussed in Boxes 1 and 2, showing mass models for the dwarf galaxy NGC\,1560 \cite{Gentile2010}. The shape of the observed rotation curve at $R\simeq4-6$ kpc is mirrored by the shape of the gas gravitational contribution. This is a common phenomenology in gas-rich dwarfs: mass models that scale up the baryonic contributions by ``brute force'' (above acceptable values of $M_\star$ and $M_{\rm gas}$) can usually reproduce the observed rotation-curve shapes without DM \cite{Hoekstra2001, Swaters2012}. Rephrasing a longstanding rule from Renzo Sancisi \cite{Sancisi2004}: ``for any feature in the baryonic mass profile of a galaxy, there is a corresponding feature in the rotation curve, and vice versa''. This is the backbone of the radial acceleration relation, discussed in the section on dynamical laws.

\subsection{Noncircular motions}

In addition to regular rotation ($V_{\rm rot}$) and turbulent motions ($\sigma_{\rm V}$), large-scale noncircular motions may exist in gas-rich galaxies. In a $\Lambda$CDM context, noncircular motions are expected in the dwarf regime for several reasons: (i) DM halos are predicted to be triaxial or prolate driving noncircular motions in the disk plane \cite{Hayashi2007}, (ii) the strong stellar feedback implemented in most hydrodynamic simulations should drive gas outflows outside the disk plane, following the path of least resistance \cite{Dutton2017, Collins2021}, and (iii) cold gas accretion from cosmic filaments may be on-going at $z=0$ in low-mass galaxies driving radial gas flows \cite{Keres2005}.

Noncircular motions in the disk plane can be studied via harmonic decompositions of the observed velocity maps \cite{Schoen1997}. In typical star-forming dwarfs (excluding starburst), \hi\ studies \cite{Gentile2005, Trachternach2008, Oh2015, Marasco2018, Hunter2019} find that planar noncircular motions range from $\sim$1$\%$ to $\sim$10$\%$ of $V_{\rm rot}$ and are generally smaller than $5-10$ km\,s$^{-1}$. This is comparable to typical uncertainties in $V_{\rm rot}$, which are usually computed considering the difference between the approaching and receding sides of the disk \cite{Swaters2009}. Thus, noncircular motions are effectively captured in the error budget. H$\alpha$ studies \cite{Spekkens2007, Simon2005} show that noncircular motions in the disk plane could be stronger than those from \hi\ studies, possibly due to the fact that ionized gas is more closely related to the star-formation sites than atomic gas (e.g., expanding bubbles and shells on small spatial scales). Importantly, observed noncircular motions are not strong enough to ``hide'' a central DM cusp \cite{Gentile2005, Trachternach2008, Oh2015}. Opposite conclusions are found analysing simulated dwarf galaxies \cite{Oman2019} because noncircular motions in simulated dwarfs appear to be stronger than in real ones \cite{Marasco2018}. Interestingly, significant radial motions of $\sim$5$-$15 km\,s$^{-1}$ (corresponding to about 10$\%$ to 40$\%$ of $V_{\rm c}$) have been detected in the disk plane of some starburst dwarfs \cite{Lelli2012a, Lelli2012b, Elson2011, Gentile2007}. In general, one cannot determine the direction of these radial motions because it is unknown which side of the disk is the nearest to the observer. Starburst dwarfs have a central \hi\ mass excess \cite{Lelli2014a, vanZee1998}, so it is likely that the radial motions are planar inflows driven by dwarf-dwarf interactions and/or cold gas accretion \cite{Lelli2012a, Lelli2014c, Ashley2017, LopezSanchez2012}.

Noncircular motions outside the disk plane are generally referred to as ``inflows'' or ``outflows''. For the latter ones, it is important to determine whether the gas will escape from the galaxy potential well (blowouts) or fall back onto the disk (galactic fountains). In typical star-forming dwarfs (excluding starbursts), there is no observational evidence at any wavelength for large-scale blowouts. Instead, there is evidence for thick gas layers extending up to $\sim$2-3 kpc above the disk plane \cite{Roychowdhury2010, Uson2003, Uson2008}. The thick gas layer seems to rotate slower than the thin disk, analogously to extra-planar gas in spiral galaxies \cite{Sancisi2008, Marasco2019}, but the presence of warps and flares complicates the kinematic analysis \cite{Kamphuis2011}. Extraplanar gas in star-forming dwarfs may be explained by supernovae-driven galactic fountains, creating a small-scale gas circulation just above the disk plane \cite{Melioli2015}.

In starburst dwarfs, one may expect to find ubiquitous and multiphase gas blowouts \cite{Dekel1986} but the observational situation is complex. Regarding cold gas ($T\lesssim10^4$ K), absorption-line observations of the Na doublet of six starburst dwarfs \cite{Schwartz2004} provided evidence for gas outflows in three cases but (i) the velocities of the outflowing gas do not clearly exceed the escape velocity and (ii) the mass of the outflowing gas cannot be easily estimated from absortion-line data. For larger samples of starburst dwarfs, emission-line \hi\ observations that directly probe the cold gas mass do not show clear evidence for gas blowouts \cite{Lelli2014a, Roychowdhury2012}. Regarding warm ionized gas ($T\simeq10^4$ K), narrow-band H$\alpha$ imaging reveals diffuse emission in $\sim$50$\%$ of starburst dwarfs, often characterized by shells and bubbles with sizes up to 2 kpc \cite{McQuinn2019}. H$\alpha$ spectroscopy shows that the velocities of the warm ionized gas are generally below the galaxy escape velocity \cite{Martin1998, vanEymeren2007, vanEymeren2009a, vanEymeren2009b} apart from a few extreme exceptions \cite{Cresci2017}. Larger outflow velocities for warm ionized gas have been reported using UV absorption lines \cite{Heckman2015, Chisholm2017}. Regarding hot gas ($T\simeq10^6-10^7$ K), several starburst dwarfs show diffuse X-ray emission \cite{Ott2005, McQuinn2018} which likely traces blowouts because the thermal velocity of the hot gas exceed the galaxy escape velocity. The observed mass of the hot gas, however, is less than 1$\%$ of the ISM mass \cite{Ott2005}.

\section{Dynamical laws}

Galaxies follow several dynamical laws, i.e., tight empirical correlations between a baryonic quantity (from the distribution of gas and stars) and a dynamic one (from the observed kinematics). In a $\Lambda$CDM context, the internal dynamics of dwarf galaxies are largely dominated by the DM halo, so it is pivotal to study the bottom-end of the dynamical laws to quantify the relation between baryons and DM. I will focus on three main dynamical laws for rotation-supported galaxies: the BTFR that applies at large radii, the CDR that applies at small radii, and the RAR that applies at each and every radius. 

\subsection{Baryonic Tully-Fisher relation}

The classic Tully-Fisher (TF) relation \cite{Tully1977} links the galaxy luminosity in optical or near-IR bands (proxy for the stellar mass) to the width of the spatially integrated \hi\ line profile (proxy for the circular velocity, after correcting for disk inclination). The classic TF relation, however, breaks down for dwarf galaxies with $V_{\rm c}\simeq100$ km\,s$^{-1}$ because their gas mass can be higher than their stellar mass \cite{McGaugh2000} (see Fig.\,1). The BTFR \cite{McGaugh2000} replaces the stellar mass with the total baryonic mass ($M_{\rm bar} = M_\star + M_{\rm gas}$), recovering a single linear relation over $\sim$6 dex in $M_{\rm bar}$ (see Fig.\,5, a).

It has been determined empirically \cite{Verheijen2001b, Lelli2019} that the scatter along the BTFR is minimized using the mean circular velocity along the flat part of the rotation curve ($V_{\rm flat}$). Other characteristic velocities, such as the velocity $V_{2.2}$ at 2.2 exponential scale lengths ($R_{\rm d}$), give BTFRs with larger scatters and shallower slopes. This is counter-intuitive because baryons are dynamically important at small radii and their gravitational contribution should approximately peak at 2.2$R_{\rm d}$ \cite{Freeman1970}, so one may expect $V_{2.2}$ to be more closely related to $M_{\rm bar}$ than $V_{\rm flat}$. This does not occurr.

The $M_{\rm bar}-V_{\rm flat}$ relation has an orthogonal intrinsic scatter of just $\sim$6 \cite{Lelli2019}. The residuals do not correlate with other galaxy properties such as scale length or surface brightness, so the scatter cannot be further decreased using a third independent variable \cite{Lelli2016b, Desmond2019}. Consequently, the relation between the specific angular momentum ($j \simeq V_{\rm flat} \times R_{\rm d}$) and the visible mass of disk galaxies \cite{Posti2018, Mancera2021} has larger scatter than the BTFR \cite{Lelli2019, Posti2020}. For circular velocities smaller than $\sim$40 km\,s$^{-1}$, a correlation between BTFR residuals and disk scale length has been reported \cite{Mancera2020} but the data are scarce with large uncertainties due to asymmetric-drift corrections and possibly disk-thickness effects \cite{Iorio2017, Bacchini2020}. TDGs \cite{Lelli2015} and UDGs \cite{Mancera2019} have been found to deviate from the BTFR but the available kinematic data are very uncertain.

The $M_{\rm bar}-V_{\rm flat}$ relation has a fitted slope close to four \cite{McGaugh2000, McGaugh2012, McGaugh2020, Lelli2019, Verheijen2001b}, so can be expressed as $M_{\rm bar} = N V_{\rm flat}^4$. By dimensional arguments, the physical units of the intercept $N$ are acceleration times $G_{\rm N}$ (with $G_{\rm N}$ being Newton's gravitational constant), so the BTFR defines a purely empirical acceleration scale:
\begin{equation}
 a_{\rm BTFR} = \dfrac{x}{N G_{\rm N}}
\end{equation}
where $x=0.80\pm0.05$ is a dimensionless parameter accounting for the disk geometry of rotating galaxies. Fixing the slope to 4, a linear fit to the data gives $a_{\rm BTFR}=(1.3 \pm 0.1) \times 10^{-10}$ m\,s$^{-1}$, where the uncertainty on $x$ dominates the error budget. The observed properties of the BTFR were predicted a-priori by MOND (see Box 2).

In a $\Lambda$CDM context, the relation between halo mass and halo velocity has a slope equal to 3 \cite{McGaugh2012}. If we assume that (i) $V_{\rm flat}$ is a proxy of the halo velocity and (ii) DM halos contain all baryons available from cosmology (i.e., we set the normalization of the theoretical $M_{\rm bar}-V_{\rm flat}$ relation using the baryon fraction from $\Lambda$CDM fits to the cosmic microwave background), we obtain the dashed line in Fig.\,5 (a). The data lies systematically to the left of this line, implying that the baryons observed in galaxies are much less than the cosmic ones. The ``missing baryons'' are commonly thought to reside in a hot gaseous halo around galaxies, which is heated by stellar feedback processes \cite{Dutton2017, Katz2018}. Thus, in a $\Lambda$CDM context, the value of $a_{\rm BTFR}$ relates to the average amount of missing baryons in galaxies. Given the difference between the predicted ($\sim$3) and observed ($\sim$4) slopes, the amount of missing baryons must systematically change with circular velocity (or halo mass), ranging from $\sim$50$\%$ in high-mass spirals down to $\sim$99$\%$ in low-mass dwarfs. Intriguingly, these baryon fractions are roughly consistent with those expected from $\Lambda$CDM abundance-matching relations \cite{DiCintio2016}. On the other hand, the small intrinsic scatter and lack of curvature in the observed BTFR are at variance with abundance-matching prescriptions \cite{Desmond2017b}. In general terms, given the small BTFR scatter, a galaxy with a given $V_{\rm f}$ (or halo mass) must ``know'' exactly how many baryons to hide in order to stay in the ``right'' location of the BTFR. This represents a conceptual fine-tuning problem for $\Lambda$CDM models of galaxy formation \cite{McGaugh2012}.

\begin{figure}
\includegraphics[width=\linewidth]{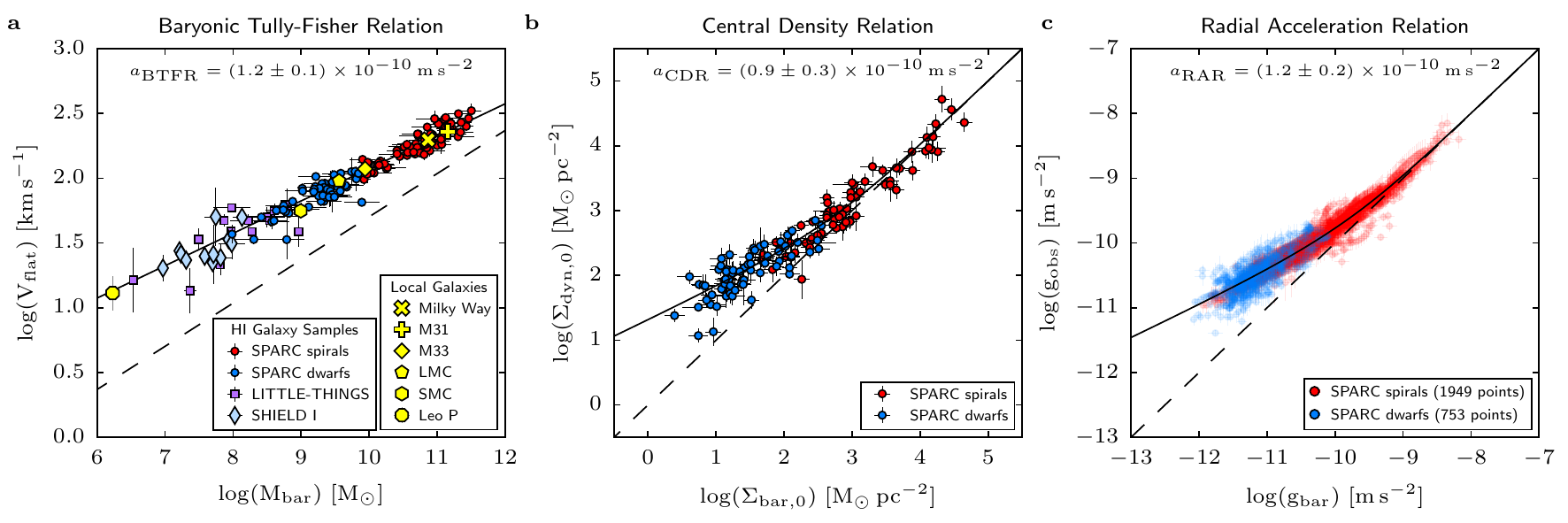}
\caption{\textbf{Dynamical laws for dwarf and spiral galaxies}. The baryonic Tully-Fisher relation (a), the central density relation (b), and the radial acceleration relation (c). In panel a, the dashed line shows the $\Lambda$CDM prediction assuming that galaxies contain all available cosmological baryons. For SPARC galaxies \cite{Lelli2019} $V_{\rm flat}$ is directly measured, while for LITTLE-THINGS galaxies \cite{Iorio2017} and SHIELD-I galaxies \cite{McNichols2016} the outermost circular velocity is used as a proxy for $V_{\rm flat}$. Yellow symbols show representative galaxies in or near the Local Group: the Milky Way (cross) \cite{McGaugh2019, Eilers2019}, M31 (plus) \cite{Chemin2009, Corbelli2010}, M33 (large diamond) \cite{Corbelli2014, Kam2017}, the LMC (pentagon) \cite{Staveley2003, vanDerMarel2006, vanDerMarel2016}, the SMC (hexagon) \cite{DiTeodoro2019}, and Leo\,P (octagon) \cite{Bernstein2014, McQuinn2015}. In panels b and c, the dashed line corresponds to the one-to-one line. In all panels, black lines show empirical fitting functions with only one free parameter: a characteristic acceleration scale (given at the top). For internal consistency, the fits are performed considering only the SPARC data (red and blue circles corresponding to spiral and dwarf galaxies, respectively). The errorbars correspond to 1$\sigma$ uncertainties.}
\label{fig:laws}
\end{figure}
\subsection{Central density relation}

The CDR is complementary to the BTFR because it involves quantities measured at small radii \cite{Lelli2016c}: the central baryonic surface density ($\Sigma_{\rm bar, 0}$) from stellar-mass maps and the central dynamical surface density ($\Sigma_{\rm dyn, 0}$) from rotation curves. The root of the CDR is a relation between the central surface brightness and the inner rotation-curve steepness \cite{Swaters2009, Garrido2005, Lelli2013, ErrozFerrer2016}, which is proportional to the square root of $\Sigma_{\rm dyn, 0}$ \cite{Lelli2014d}. The CDR exploits an exact analytic formula for self-gravitating disks \cite{Toomre1963}:
\begin{equation}\label{eq:Toomre}
 \Sigma_{\rm dyn, 0} = \dfrac{1 + q_{0}}{2\pi G_{\rm N}} \int_{0}^{\infty} \dfrac{V_{\rm c}^{2}(R)}{R^{2}}dR,
\end{equation}
where $q_0$ is the intrinsic axial ratio of the disk \cite{Lelli2016c}. The CDR formally holds for $R\rightarrow0$ but in practice $\Sigma_{\rm dyn, 0}$ is measured within one resolution element of the kinematic data (typically a few hundred parsecs). The mean baryonic surface density is then measured within the same aperture. At these small radii, the stellar surface density is much larger than the gas surface density, so the gas contribution can be neglected in measuring $\Sigma_{\rm bar, 0}$. A relation similar to the CDR exists between the central surface brightness and the central dynamical surface density measured from the vertical velocity dispersion of stellar disks in face-on spirals \cite{Swaters2014}.

The CDR is shown in Figure\,\ref{fig:laws} (b). Differently from the BTFR, the relation is non-linear. For surface densities above $\sim$500 M$_\odot$ pc$^{-2}$, the data are along the one-to-one relation indicating that baryons dominate the inner dynamics. At surface densities below $\sim$500 M$_\odot$ pc$^{-2}$, the data systematically deviate from unity indicating that the DM effect is already significant in the inner regions. Fixing the top-end slope to one, the CDR can be fitted by a double power law with three free parameters: the bottom-end slope and two characteristic surface densities, one setting the transition point from the unity line and one setting the overall normalization \cite{Lelli2016b}. The two characteristic surface densities have approximately the same value \cite{Lelli2016b}, so we will refer to both of them as $\Sigma_{\rm CDR}$ for simplicity. The bottom-end slope of the CDR is close to $0.5-0.6$.

By dimensional arguments, surface densities can be expressed in units of acceleration over $G_{\rm N}$, so the characteristic surface density $\Sigma_{\rm CDR}$ implies the existence of another acceleration scale:
\begin{equation}\label{eq:CDR}
 a_{\rm CDR} = 2\pi G_{\rm N} \Sigma_{\rm CDR},
\end{equation}
which is conceptually different from $a_{\rm BTFR}$ because it sets the transition from baryon-to-DM domination in the inner galaxy regions. In a $\Lambda$CDM context, this characteristic scale is presumably related to the angular momentum of baryonic gas, i.e., its ability to collapse to the bottom of the DM halo and form stars, reaching high stellar surface densities. In a MOND context, the CDR is naturally explained \cite{Milgrom2016} and can be fitted by a function with a single free parameter: $\Sigma_{\rm CDR}$ or equivalently $a_{\rm CDR}$. Irrespective of the physical meaning of MOND, the black line in Fig.\,5\,(b) provides a practical, effective description of the CDR shape. Using this function, we find $a_{\rm CDR} = (0.9\pm0.3)\times10^{-10}$ m\,s$^{-2}$ where the error budget is dominated by systematic uncertainties in $q_0$ \cite{Lelli2016c}.

The CDR has an observed scatter of $\sim$0.2 dex ($\sim$45\%) that is largely driven by observational uncertainties. The residuals show no correlation with other galaxy properties like stellar mass or scale length. This is unexpected because Newton's shell theorem does not hold in disk galaxies, so the mass outside a given radius contribute to the gravitational field within that radius. In fact, Eq.\,\ref{eq:Toomre} formally integrates $V^2_{\rm c}/R^2$ from zero to infinity. Thus, in a Newtonian context, one may expect the residuals around $\Sigma_{\rm dyn, 0}$ to correlate with galaxy mass or size. This is not observed.

To date there has been no attempt to reproduce the CDR in a $\Lambda$CDM context. However, the ``unexpected diversity'' of the dwarf galaxy rotation curves \cite{Oman2015} suggests that the existence of this relation is not ``natural'': one may expect $\Sigma_{\rm dyn, 0}$ to correlate more closely with stellar mass (or halo mass) than stellar surface density, but the opposite is observed \cite{Lelli2016b}.

\subsection{Radial acceleration relation}

The RAR \cite{McGaugh2016, Lelli2017} relates the observed centripetal acceleration from rotation curves ($g_{\rm obs} = V_{\rm c}^2/R$) to the baryonic gravitational field ($g_{\rm bar} = -\nabla \Phi_{\rm bar}$), which can be computed by solving the Poisson's equation for the observed distribution of stars and gas ($\nabla^2 \Phi_{\rm bar} = 4\pi G_{\rm N} \rho_{\rm bar}$). The RAR applies at each radius in rotation-supported galaxies, so it is a spatially resolved relation that embeds the BTFR for $R\rightarrow \infty$ and the CDR for $R\rightarrow 0$.

The RAR is non-linear (Fig.\,5, c) and displays two explicit acceleration scales: one setting the transition value from the unity line and one setting the overall normalization. The two acceleration scales have approximately the same value \cite{Lelli2017}, similarly to the surface-density scales in the CDR \cite{Lelli2016c}, so we will refer to them as $a_{\rm RAR}$ for simplicity. Above $a_{\rm RAR}$, we have $g_{\rm obs}=g_{\rm bar}$ because the data largely come from the inner baryon-dominated regions of spiral galaxies. Below $a_{\rm RAR}$, we have $g_{\rm obs}>g_{\rm bar}$ because the data come either from the outer DM-dominated regions of spiral galaxies or from nearly every radii of dwarf galaxies. The fact that spiral and dwarf galaxies smoothly connects in the low-acceleration part of the RAR is highly non-trivial: the amount of DM in the outer parts of spirals is somehow comparable to that in the inner parts of dwarfs. The residuals around the RAR show no relation with other galaxy properties \cite{Lelli2017} and the observed scatter is consistent with observational uncertainties \cite{McGaugh2016, Lelli2017, Li2018}.

In a $\Lambda$CDM context, there have been several attempts to reproduce the RAR but the results are mixed. Some hydrodynamic simulations of galaxy formation reproduce the RAR with the observed value of $a_{\rm RAR}$ \cite{Keller2017, Garaldi2018}, other simulations give a relation with too high value of $a_{\rm RAR}$ \cite{Ludlow2017}, while other ones display no acceleration scale at all \cite{Wu2015, Tenneti2018}. This illustrates that, contrary to some claims \cite{Navarro2017}, the RAR shape is not a ``natural'' outcome of $\Lambda$CDM cosmology but depends on how different models treat the complex interplay between baryons and DM during the formation and evolution of galaxies. The most severe challenge for $\Lambda$CDM models, however, is not the RAR shape but the RAR scatter, which appears to be significantly smaller than predicted by fiducial abundance-matching prescriptions \cite{Desmond2017}.

The shape of the RAR is similar to that of the CDR but the two functional forms are different. In a MOND framework (see Box\,2), the RAR shape is given by the derivative of the CDR shape \cite{Milgrom2016}. Fitting a CDR-motivated function to the RAR data, we obtain $a_{\rm RAR} = (1.2\pm0.2) \times 10^{-10}$ m\,s$^{-2}$ where the error budget is dominated by systematic uncertainties in the conversion from stellar light to stellar mass. Other empirical fitting functions provide similar values of $a_{\rm RAR}$ \cite{McGaugh2016, Lelli2017}. In a DM context, it is remarkable that $a_{\rm BTFR} \simeq a_{\rm CDR} \simeq a_{\rm RAR}$ within the uncertainties because these acceleration scales have distinct physical meanings: $a_{\rm BTFR}$ sets the BTFR intercept and is related to the average amount of missing baryons, $a_{\rm CDR}$ sets the transition from baryon to DM dominated galaxies in the inner regions and is presumably related to the baryonic angular momentum, whereas $a_{\rm RAR}$ sets the transition from baryon-to-DM domination inside each galaxy and is related to the complex history of gas, stellar, and DM build-up at each radius. To make an explicit example: for $g_{\rm bar} \ll a_{\rm RAR}$ we have $g_{\rm obs} = \sqrt{a_{\rm RAR}g_{\rm bar} }$ that gives $V^4_{\rm flat} = a_{\rm RAR} G_{\rm N} M_{\rm bar}$ at large radii \cite{Lelli2017}. This is equivalent to the BTFR but it is unclear why $a_{\rm RAR}$ (related to the internal mass distribution) should have the same numerical value as $a_{\rm BTFR}$ (related to the average amount of missing baryons). In a MOND context, instead, one must necessarily have $a_{\rm BTFR} = a_{\rm CDR} = a_{\rm RAR}$ because these acceleration scales subtend the existence of a new constant of Nature: $a_0$ \cite{Milgrom1983a, Milgrom1983b, Milgrom1983c}. Thus it is of outmost importance to understand (i) why these acceleration scales exist, (ii) whether they have exactly the same value, (iii) which physical processes set their values, and (iv) whether they evolve or not with cosmic time.

\section{Outlook}

\subsection{Next-generation kinematic surveys}

The outstanding progress I reviewed here has been possible thanks to \hi\ and H$\alpha$ kinematic data accumulated over the past 40 years by several generations of astronomers. This has allowed studying gas dynamics in all known types of gas-rich dwarfs, spanning $\sim$4 dex in stellar mass. The available kinematic samples, however, remain relatively small (a few hundreds galaxies) and rather heterogeneous in nature. These considerations are important when studying dynamical laws and their intrinsic scatters, which are a key testbed for both $\Lambda$CDM models \cite{Desmond2017b, Desmond2017, Dutton2017, Garaldi2018} and alternative theories \cite{McGaugh2012, Lelli2017, Li2018, Chae2020}. $\Lambda$CDM models, indeed, generally predict that there must be non-negligible intrinsic scatter around the dynamical laws, driven by the diverse evolutionary paths followed by galaxies and their DM halos \cite{Desmond2017b, Desmond2017}. This intrinsic scatter is expected to become increasingly large towards the least massive galaxies \cite{Dutton2017, Garaldi2018}. On the other hand, MOND predicts tiny or null intrinsic scatter around these dynamical laws, depending on the precise theory formulation \cite{Famaey2012} and on the galaxy environment \cite{Chae2020}. 
Obtaining large and homogeneous kinematic samples, probing the faintest dwarfs, is the next observational challenge.

The future looks bright. The Square Kilometer Array (SKA) observatory and its pathfinders are expected to provide spatially resolved \hi\ observations for several thousand galaxies at $z\simeq0$ \cite{Koribalski2020, Duffy2012, deBlok2015}, possibly detecting dwarf galaxies with \hi\ masses around $\sim$10$^5$ M$_\odot$ or less \cite{Adams2018}. These next-generation \hi\ surveys will be complementary to planned optical and near-IR wide-field surveys, probing the stellar component of dwarf galaxies, for example with the upcoming Euclid space mission \cite{Euclid} and the Vera C. Rubin observatory \cite{LSST}. In the coming years, therefore, we will gather large, homogeneous samples of dwarf galaxies, enlarging the sizes of the current samples by orders of magnitude. This will allow us to measure the intrinsic scatter in the dynamical laws of galaxies with high accuracy and investigate their possible environmental dependence in a robust way. In this context, it will be beneficial to have clear a-priori predictions from $\Lambda$CDM simulations tailored to the planned surveys, before the observations are actually taken.

\subsection{Next-generation modeling techniques}

While statistical studies of dwarf galaxies will be crucial in the near future, detailed high-resolution studies of individual objects will also be game-changing. For example, the SKA observatory is expected to provide \hi\ data with unprecedented spatial resolutions and sensitivities, probing gas dynamics in significant samples of dwarf galaxies at small spatial scales (below 10 pc) and column densities ($\sim$10$^{18}$ cm$^{-2}$). In this context, it will be necessary to extend existing 3D modeling techniques \cite{Jozsa2007, Kamphuis2015, DiTeodoro2015} to include complex small-scale noncircular motions. Moreover, in the lowest mass galaxies (with $V_{\rm rot}\lesssim20$ km\,s$^{-1}$), accurate modeling of pressure support and disk thickness will become important \cite{Iorio2017, McQuinn2021}. A promising approach is to simultaneously model the midplane gravitational potential in the radial direction together with that in the vertical direction assuming hydrostatic equilibrium \cite{Bacchini2020}.

From the galaxy formation perspective, it will be important to compare observed galaxies with simulated ones in a robust way. The most promising approach is building mock observations of simulated galaxies, taking observational effects into account \cite{Read2016b, Verbeke2017, Marasco2018, Oman2019}. The goal will be to assess whether $\Lambda$CDM models are able to form realistic dwarf galaxies by reproducing diverse (yet interrelated) aspects of their gas kinematics, such as the rotation-curve shape, the DM content, the strength of noncircular motions, the gas velocity dispersion, and the occurrence of extraplanar gas. Most existing models are often able to reproduce a subset of them (e.g., rotation curve shapes using strong stellar feedback) but fail on others (e.g., too strong noncircular motions \cite{Marasco2018}). This does not occur by chance. For example, strong stellar feedback is commonly implemented to solve the cusp-vs-core and too-big-too-fail problems (see Box\,1), but may then generate too strong noncircular motions and/or too massive gas outflows. Again, it will be desiderable to build samples of mock-observed galaxies before the actual observations are taken, so one can have clear and unambiguous predictions from different galaxy formation models. This is the only way to adhere to the scientific method \cite{Merritt2017} and push forward our understanding of galaxy formation and evolution.

\section*{Acknowledgements}

Writing this review would have not been possible without the many enlightening conversations with many colleagues over the years. In particular I thank Alice Concas, Edvige Corbelli, Filippo Fraternali, Stacy McGaugh, James Schombert, and Paolo Tozzi for their precious comments on a first draft of this manuscript.

\bibliography{GasDynamics}

\end{document}